\documentclass[prl,showpacs,preprintnumbers,twocolumn]{revtex4}
\usepackage{epsfig}
\usepackage{amssymb,amsmath,amsthm}

\usepackage[draft]{hyperref} 

\begin{document}

\title{Optical wave turbulence and the condensation of light}


\author{Umberto Bortolozzo$^{1}$, Jason Laurie$^{2,*}$, Sergey Nazarenko$^{2}$
and
Stefania Residori$^{1}$}

\address{$^{1}$INLN, Universit\'e de Nice Sophia-Antipolis, CNRS, 1361 route
des Lucioles 06560 Valbonne, France\\
$^{2}$Mathematics Institute, University of Warwick, Coventry CV4
7AL,
United Kingdom\\
$^{*}$Corresponding author: j.p.laurie@warwick.ac.uk}

\begin{abstract}
In an optical experiment, we report a wave turbulence regime that, starting with
weakly nonlinear waves with randomized phases, shows an inverse cascade of
photons towards the lowest wavenumbers. We show that the cascade is induced by a
six-wave resonant interaction process and is characterized by increasing
nonlinearity. At low wavenumbers the nonlinearity becomes strong and leads to
modulational instability developing into solitons, whose number is decreasing
further along the beam.
\end{abstract}


\maketitle 

The idea to create the state of optical wave turbulence (OWT) has excited the
minds of scientists for over thirty years, and it was a subject of a rather
large number of theoretical papers \cite{OT,OT2,OT3,OT4,NO}. Indeed there are
some far reaching fluid analogies in the dynamics of nonlinear light, for
example vortex-like solutions \cite{OV, OV2}, shock waves \cite{Fleischer} and
weakly interacting random waves whose dynamics and statistics has similarities
to the system of random waves on water surface \cite{zf,zlf}. OWT was
theoretically predicted to exhibit dual cascade properties similar to 2D fluid
turbulence, namely the energy cascading directly, from low to high frequencies,
and the photons cascading inversely, toward the low energy states
\cite{zlf,OT,OT2,OT3,OT4}.
When the nonlinearity is small, OWT can be
described by the theory of {\em weak turbulence} (WT) \cite{zlf} which possesses
classical attributes of the general turbulence theory, particularly predictions
of the Kolmogorov-like cascade states, which in the WT context are called
Kolmogorov-Zakharov (KZ) spectra. It appears that OWT has two KZ states: one
describing the direct energy cascade from large to small scales, and the second
one - an inverse cascade of wave action toward larger scales.
The inverse cascade is particularly interesting because in the optics context it
means condensation of photons.
Furthermore, it was theoretically predicted, and numerically observed in some WT systems, that in the course of the inverse
cascade the nonlinearity will grow, which will eventually lead to the breakdown of
the WT description at low wavenumbers and to the formation
of coherent structures  \cite{OT,OT3,NO,MMT97,CMMT01,DPZ04}. In optics, these can be solitons or collapses for focusing nonlinearity
or a quasi-uniform condensate and vortices for the defocusing case.

The final thermalized state was studied extensively theoretically in various
settings for non-integrable Hamiltonian systems starting with the pioneering
paper by Zakharov {\em et al} \cite{ZPSY88}, see also
\cite{RN01,RN03,BKKMMP09,ET06,JJ00,RCK00,JTZ00,PPM08,PLJP06,PY85}.
The final state with a single soliton and small-scale noise was interpreted as a
statistical attractor, and an analogy was pointed
out to the over-saturated vapor system, where the solitons are similar to
droplets and the random waves are like molecules \cite{PY85}. There,
small droplets evaporate while the big ones gain size from the free molecules, resulting in the decrease in the number of droplets.  In our work we put emphasis on turbulence, i.e. on a transient non-equilibrium process leading to thermalization rather than the thermal equilibrium itself.

We report the first experimental evidence of OWT: starting with
weakly nonlinear incoherent waves we observe an inverse cascade of the photons
to lower wavenumbers and growth of the nonlinearity leading to the formation of
multiple coherent solitons, which further merge into a single strong soliton.
This corresponds to Bose-Einstein condensation (BEC) of photons, which is an
optical analogue of atomic BEC reported in \cite{bec, bec2}.

The motion of photons to different energy levels (and toward the
lowest one corresponding to condensation) can be achieved via nonlinear
interactions of photons, provided for example by the Kerr effect.
The problem, however, is that the nonlinearity is usually very weak and it is an
experimental challenge
to make it overpower the dissipation. In our view this was the main obstacle
with the photon condensation setup in a 2D Fabry-Perot cavity
theoretically suggested in \cite{chiao}.

The key feature in our setup is that we traded one spatial dimension for a
``time axis", namely we considered a time independent 2D light field where the
principal direction of the light propagation plays the role of ``time".
This allowed us to use a nematic liquid crystal, which provides a high level and
tunable optical nonlinearity \cite{Zeldovich,Khoo}. The slow relaxation time of
the reorientational dynamics is not a restriction in our setup because our
system is steady in time. Similar experiments were first reported in
\cite{Libchaber}, where it was shown that a beam propagating inside a nematic
layer undergoes a strong self-focusing effect followed by filamentation,
soliton formation, when the light intensity is increased. More recently, a
renewed
interest in the same setup has led to further studies on the optical solitons
and modulational instability regimes \cite{Assanto,Peccianti,Conti}. However,
all these experiments have in common a large value of the input intensity, of
the order of $10^3$ $W/cm^2$, therefore the optical nonlinearity is rather high
and, as soon as the beam enter the liquid crystal, the soliton or modulational
instability regimes appear immediately, bypassing the WT regime.

In our experiment, we fix a much lower input intensity, of the order of
$10^{-1}$ $W/cm^2$, and, for the first time, we show that, before the
modulational instability and the soliton formation, corresponding to the
breakdown of the weak nonlinearity, there is a WT regime characterized by an
inverse cascade of the photon number. Besides the weak nonlinearity, to allow
for the wave-mixing and the inverse cascade development, we also need to prepare
a proper initial condition: we start with weakly nonlinear incoherent waves,
characterized by high wavenumbers and random phases.

The 1D system is rather different from the 2D and 3D systems
considered theoretically before, and in this paper, for the first time, we present
a theory of 1D OWT, which involves coexisting random waves
(interacting via a six-wave process) and solitons (to which the random waves
condense). We complement our study with the numerical simulation
of the dynamical equation, and we observe that the theory, numerics and
experiment are consistent and complementary with each other, giving a rather
complete description of the 1D OWT phenomenon.

\begin{figure}[htbp]
\centerline{\includegraphics[width=8.3 cm]{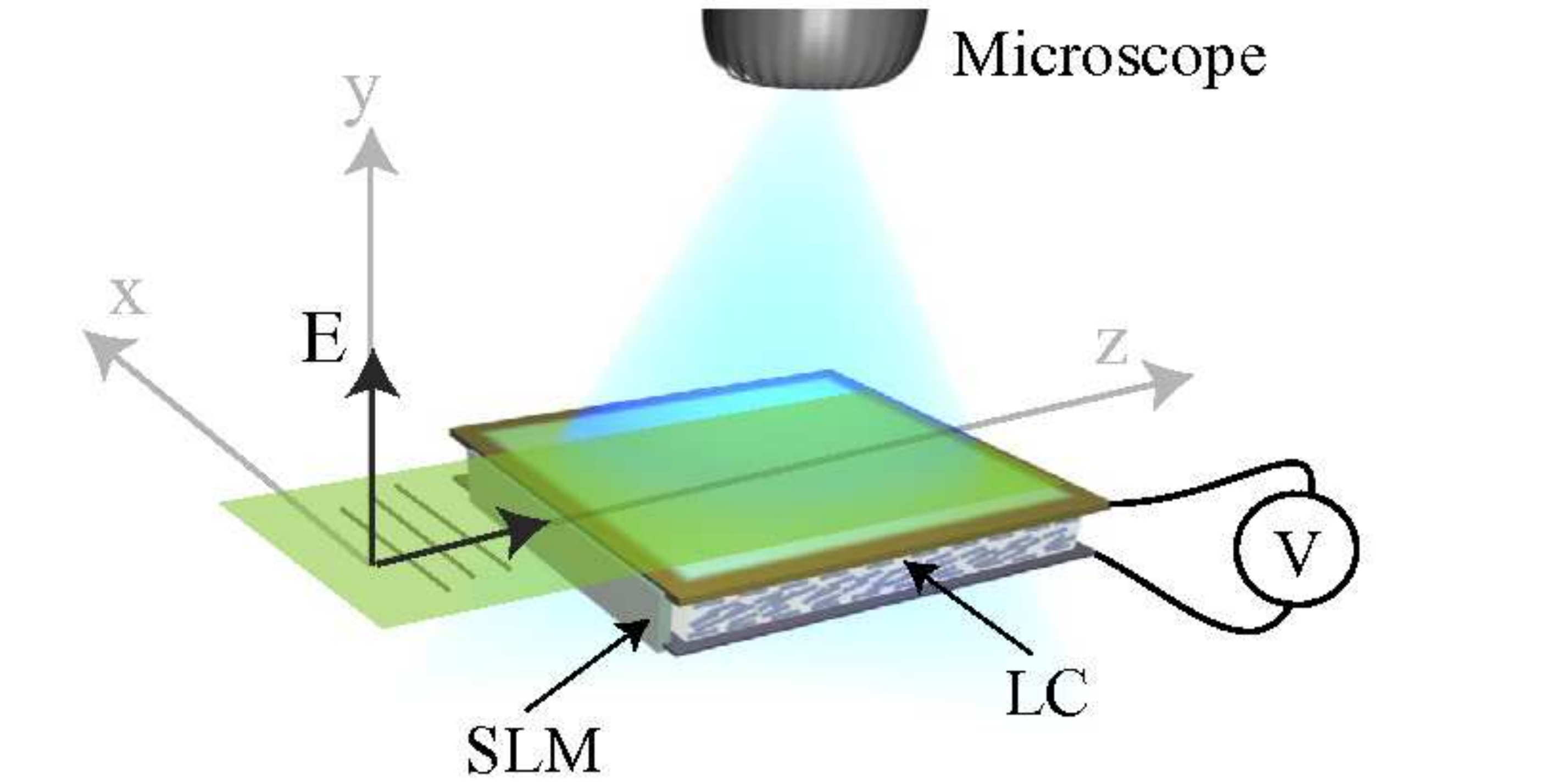}}
\caption{(color online). Experimental setup: a laminar shaped input beam
propagates inside the liquid crystal (LC) layer; random space modulations are
imposed at the
entrance of the cell by means of a spatial light modulator (SLM).}
\label{fig:setup}
\end{figure}

The experimental apparatus is shown in Fig.~\ref{fig:setup}. It consists of a
laminar shaped beam propagating inside a nematic liquid crystal (LC) layer.
Before entering the cell, the beam passes through a spatial light modulator
(SLM) that, through suitable intensity masks, modulates the input profile for
injecting random phased fields with large wavenumbers.
The LC cell is made by sandwiching a nematic layer (E48) of thickness $d=50$
$\mu m$ between two $20 \times 30$ $mm^2$ glass windows. On the interior, the
glass walls are coated with Indium-Tin-Oxide (ITO) transparent electrodes. We
have pre-treated the ITO surfaces in order to align all the molecules parallel
to the confining walls (nematic director along $\hat z$).
The LC layer behaves as a positive uniaxial medium, with $n_e =1.7$ the
extraordinary and $n_o=1.5$, the ordinary refractive indices \cite{Khoo}. LC
molecules tend to turn more along the applied field and the refractive index $n(
\theta )$ follows the distribution of the tilt angle.  We apply a $1$ $kHz$
electric field with rms voltage $V_0=2.5$ $V$ so that the molecular director is
preset to an average tilt $\theta$. When a linearly polarized beam is injected
into the cell, the LC molecules reorient towards the direction of the input
polarization. The input light comes from a diode pumped solid state laser,
$\lambda =473$ $nm$, linearly polarized along $\hat y$ and shaped as a thin
laminar Gaussian beam of $30$ $\mu m$ thickness, $1.8$ $mm$ width. The total
power was $200$ $\mu W$, corresponding to an input intensity of $370$ $mW/cm^2$,
which is low enough to ensure the weakly nonlinear regime. The beam evolution
inside the cell is monitored with an optical microscope and a CCD camera.

Theoretically, the beam evolution is described by a propagation equation for the
input beam coupled to a relaxation equation for the LC dynamics

\begin{eqnarray}
2iq {\partial \psi \over \partial z}+ {\partial^2 \psi \over \partial x^2}+
k_0^2n_a^2 a \psi &=& 0,
\label{prop}
\\
{\partial^2 a \over \partial x^2} - {1 \over l_\xi^2} a + {\varepsilon_0 n_a^2
\over 4 K} | \psi |^2&=&0,
\label{dyn}
\end{eqnarray}
where $\psi(x,z)$ is the complex amplitude of the input beam propagating along
``time axis" $\hat z$,
$x$ the coordinate across the beam, $a(x,z)$ the liquid crystal reorientation
angle,
$n_a=n_e-n_o$ the birefringence of the LC, $k_0$ the optical wavenumber,
$\varepsilon_0$ the vacuum permittivity and $l_\xi = \left(\pi K / 2 \Delta
\varepsilon\right)^{1/2} (d/ V_0)$ the electrical coherence length of the LC
\cite{DeGennes}, with $K$ the elastic constant,
$q^2=k_0^2\left(n_o^2+n_a^2/2\right)$ and $\Delta \varepsilon$ the dielectric
anisotropy.
Note that $l_\xi$ fixes the typical dissipation scale, limiting the extent of
the inertial range in which the WT cascade develops. In other contexts, see e.g.
\cite{Assanto,Peccianti,Conti}, such a spatial diffusion of the molecular
deformation has been denoted as a nonlocal effect. In our experiment, for
$V_0=2.5$ $V$ we have $l_\xi =9$ $\mu m$.
The experimental setup corresponds to the long-wave regime, $kl_\xi \ll 1$, where $k$ is the typical wavenumber inside the range of the inverse cascade.  Therefore, in this paper we consider the limit $kl_\xi \ll 1$ of Eqs. (\ref{prop}) and (\ref{dyn}), which enables us to form a single equation for the input beam
\cite{us}:


 \begin{equation}
2iq {\partial \psi \over \partial z}= -{\partial^2 \psi \over \partial x^2}-
{\varepsilon_0 n_a^4 l_\xi^2 k_0^2 \over 4 K} \left ( \psi |\psi |^2 + l_\xi^2
\psi {\partial^2 | \psi |^2 \over \partial x^2} \right )=\frac{\delta
H}{\delta
\psi^*}.
\label{eq:Equation}
\end{equation}

Eq. (\ref{eq:Equation}) conserves the energy

\begin{equation}
H= \int {\left |{\partial \psi \over \partial x} \right |^2-{\varepsilon_0
n_a^4
l_\xi^2 k^2_0 \over 8 K} \left [ |\psi |^4-
l_\xi^2  \left ( {\partial |\psi |^2 \over \partial x}
\right )^2  \right ] }dx,
\label{eq:Hamiltonian}
\end{equation}
and the number of photons, $N= \int  |\psi |^2 \, dx.$
In the limit ${\varepsilon_0 n_a^4 l_\xi^2 k_0^2 / 2 K} \to 0$, Eq.
(\ref{eq:Equation}) becomes
the linear Schr\"odinger equation and has linear  wave solutions
$\psi(x,z) \sim a_k \exp(-i \omega_k z + ikx)$ with ``frequencies"
$\omega = k^2$ and constant complex amplitudes $a_k$.
For weak nonlinearity the amplitude $a_k$ become weakly dependent on ``time"
$z$.
Note that the leading nonlinear term cannot generate a cascade in $k$ because it
corresponds to an {\em integrable} 1D Nonlinear  Schr\"odinger equation. Thus,
we retain the sub-leading term and apply the WT theory, which allows us to
describe weakly nonlinear waves with random phases of $a_k$.
To apply the WT theory, we also need to verify that the linear dynamics
dominate in the system. The ratio of the linear term and the leading nonlinear
term of Eq. (\ref{eq:Equation}) is

\begin{equation}
J=\frac{4K k^2 }{ \epsilon_0 n_a^4 k_0^2 l_\xi^2I},
\end{equation}
where $I$ is the input intensity. For $I=370$ $mW/cm^2$ we have $J \simeq 100$,
which ensures the validity of the
WT approach.

For the wave action spectrum $n_k= \langle a_k a_k^* \rangle$
(the averaging is over the random phases), the WT approach yields to the
following kinetic equation \cite{us},

\begin{eqnarray}
\label{eq:kinetic}
{\partial n_k \over \partial z} &=&A \int
f_{k12345} \; \delta (k+ k_1+ k_2- k_3- k_4- k_5)
\nonumber
\\
 && \times\delta (\omega_k+\omega_1+\omega_2-\omega_3-\omega_4-\omega_5)   d
k_1d k_2
d  k_3 d  k_4 d  k_5,
\end{eqnarray}
with $f_{k12345} =n_kn_1n_2n_3n_4n_5({1 \over n_k} + {1 \over n_1}+{1 \over
n_2}-{1 \over n_3}-{1 \over n_4}-{1 \over n_5})$ and $A={9 \pi k^8_0
n_a^{16}l_{\xi}^{12}\epsilon_0^4 / 2048 K^4} $.

The kinetic equation has important exact power law solutions
$n_k = C|k|^{-\nu}$ where $C$ and $\nu$ are constants.
In particular,  $\nu=0$ and $\nu=2$ correspond to thermodynamic equilibria with
equipartitions of the particle density and the energy respectively.
In addition, there are KZ power law solutions corresponding to a direct energy
cascade from low to high $k$'s,

\begin{equation}
n_k =C|k|^{-1},
\end{equation}
and an inverse wave action cascade from high to low $k$'s,

\begin{equation}\label{eq:n_k}
n_k =C|k|^{-3/5}.
\end{equation}

Note, that a kinetic equation can be derived directly from Eqs. (\ref{prop}) and (\ref{dyn}) without the assumption of $kl_\xi \ll 1$.  We will consider the general case, including the short-wave limit of Eqs. (\ref{prop}) and (\ref{dyn}) in an extended publication \cite{us}.

Here, we concentrate on the inverse cascade. The inverse cascade spectrum is of a finite capacity type, in a sense that
only a finite amount of the cascading invariant (wave action in this case) is
needed to fill the infinite inertial range. (Indeed, the integral of $n_k \sim
k^{-3/5}$ converges at $k=0$). In these cases the turbulent systems have a
long transient (on its way to the final thermal equilibrium state) in which the
scaling is of the KZ type. This is because the initial condition serves as a
huge reservoir of the cascading invariant. Note that the situation here is not
specific for WT only and it is valid generally for turbulence. For example,
it is valid for Navier-Stokes turbulence, i.e. the Kolmogorov-Obukhov spectrum,
which is also finite capacity.

To set up the inverse cascade, we have injected photons at small spatial scales with random
phases.
This is made by creating through the SLM a random distribution of diffusing
spots with the average size $\sim 35$ $\mu m$.
The numerical initial condition is more idealized and strictly localized at a
small-scale range: we excite five wavenumbers with constant amplitude around
$|k_f|\sim 1.5\times 10^2$ $mm^{-1}$, with the phase of $\psi_k $ being random
and independent at each $k$.  Moreover,
we apply a Gaussian filter in physical space to achieve a beam profile
comparable to
that of the experiment.  Applicability of the WT approach for the numerical
simulation is verified by the
calculation of parameter $J$, which agrees with the
experiment and is of the order of $J\simeq 100$.
Experimentally, we measure the light intensity $I(x,z)=|\psi|^2$ and not the
phases of $\psi$ and, therefore, the spectrum $n_k$ is not
directly accessible. Instead, we measure the spectrum of intensity $N(k,z) =
|I_k(z)|^2$. The scaling for $N_k$ in the inverse cascade
state is easy to obtain from the wave action spectrum (\ref{eq:n_k}) and the
random phase
condition; this gives

\begin{equation}\label{eq:N_k}
N_k \sim |k|^{-1/5}.
\end{equation}

\begin{figure}[htbp]
\centerline{\includegraphics[width=8.3 cm]{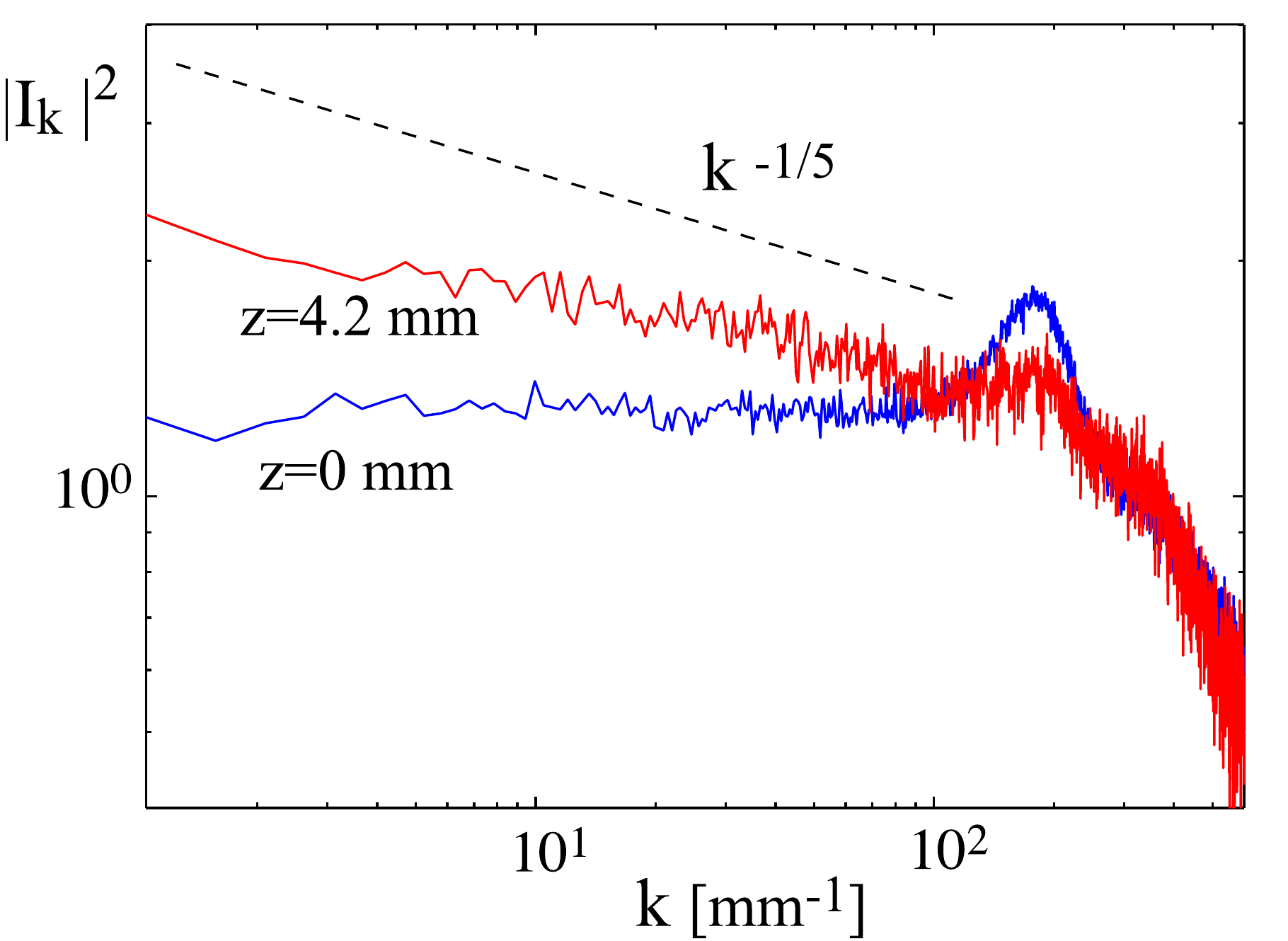}}
\caption{(color online). Experimental spectrum of the light intensity,
$N_k=|I_k|^2$ at two different distances $z$. }
\label{fig:inverse-exp}
\end{figure}

\begin{figure}[htbp]
\centerline{\includegraphics[width=8.3 cm]{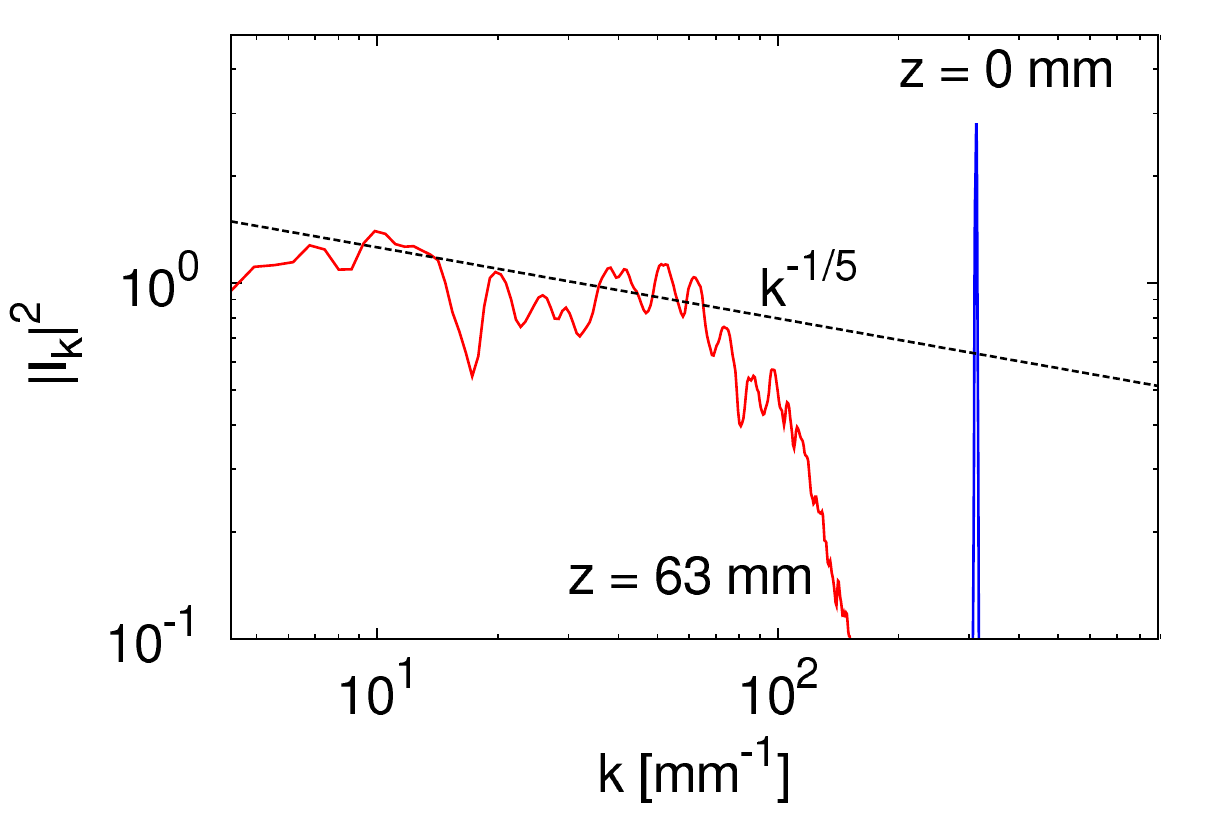}}
\caption{(color online). Numerical spectrum of the light intensity,
$N_k=|I_k|^2$ at two different distances $z$. Averaging is taken over a small
finite time window and over ten realizations.}
\label{fig:inverse-num}
\end{figure}

Experimental and numerical spectra of the light intensity are shown in
Figs.~\ref{fig:inverse-exp} and \ref{fig:inverse-num}
respectively. In both cases one can see an inverse cascade excitation of the
lower $k$ states, and good agreement
with the WT prediction of the intensity spectrum (\ref{eq:N_k}).

Closeness of Eq. (\ref{eq:Equation}) to integrability means that we should
expect not only random waves but also soliton-like coherent structures.
In the inverse cascade setup the solitons appear naturally. Indeed, the WT
description (Eq. (\ref{eq:kinetic})) breaks down when the inverse cascade
reaches some low $k$'s \cite{us}.
Modulational instability develops at these scales, which results in
filamentation of light and its condensation into coherent structures - solitons.

Verification of the long-wave limit $kl_\xi \ll 1$ and deviation from integrability is checked by considering the ratio of the two nonlinear terms in Eq. (\ref{eq:Equation}), which is estimated in Fourier space as $R \sim k^2l_\xi^2$.  We find from Figs.~\ref{fig:inverse-exp} and \ref{fig:inverse-num} that the inverse cascade is approximately in the region $k \sim 10^4 - 10^5$  $m^{-1}$ giving an estimation of $R \sim 10^{-2} - 1$.  Note, that if $R$ is too small, then we are close to a purely integrable system, which would be dominated by solitons and lack cascade dynamics, and if too high then our model breaks down.

\begin{figure}[htbp]
\centerline{
\includegraphics[width=8.3cm]{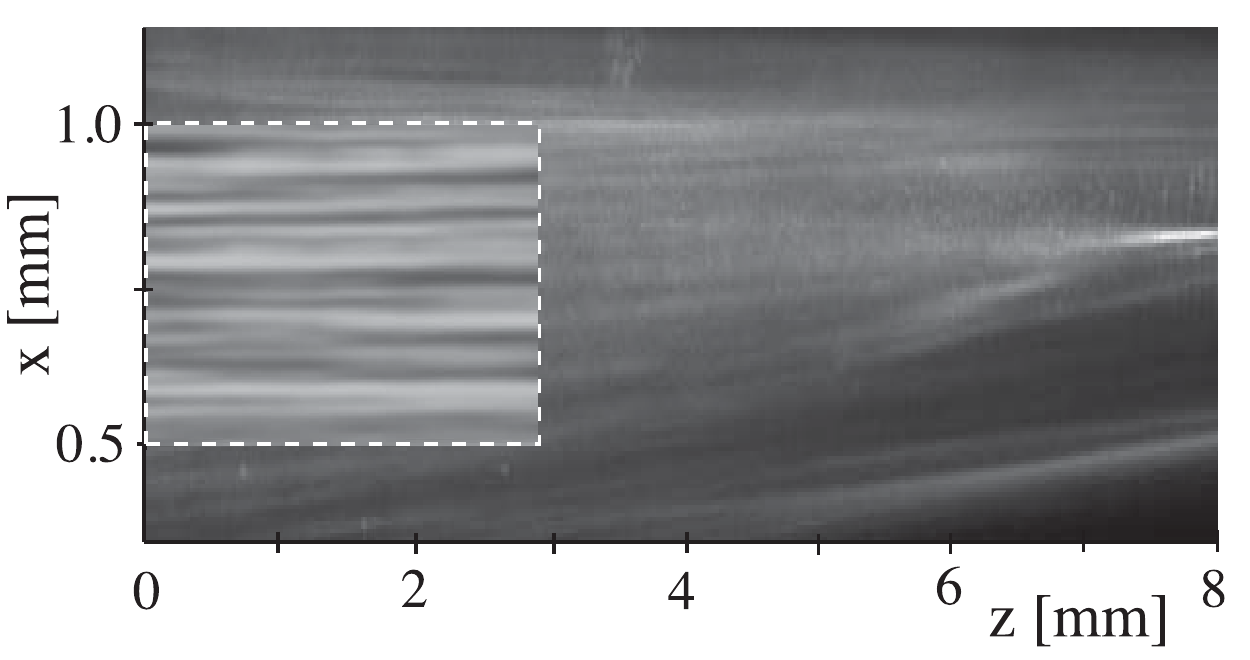}}
\caption{Experimental results for intensity distribution $I(x,z)$. Area marked
by the dashed line is shown at a higher
resolution (using a larger magnification objectif).}
\label{fig:xz-exp}
\end{figure}

\begin{figure}[htbp]
\centerline{
\includegraphics[width=8.3cm]{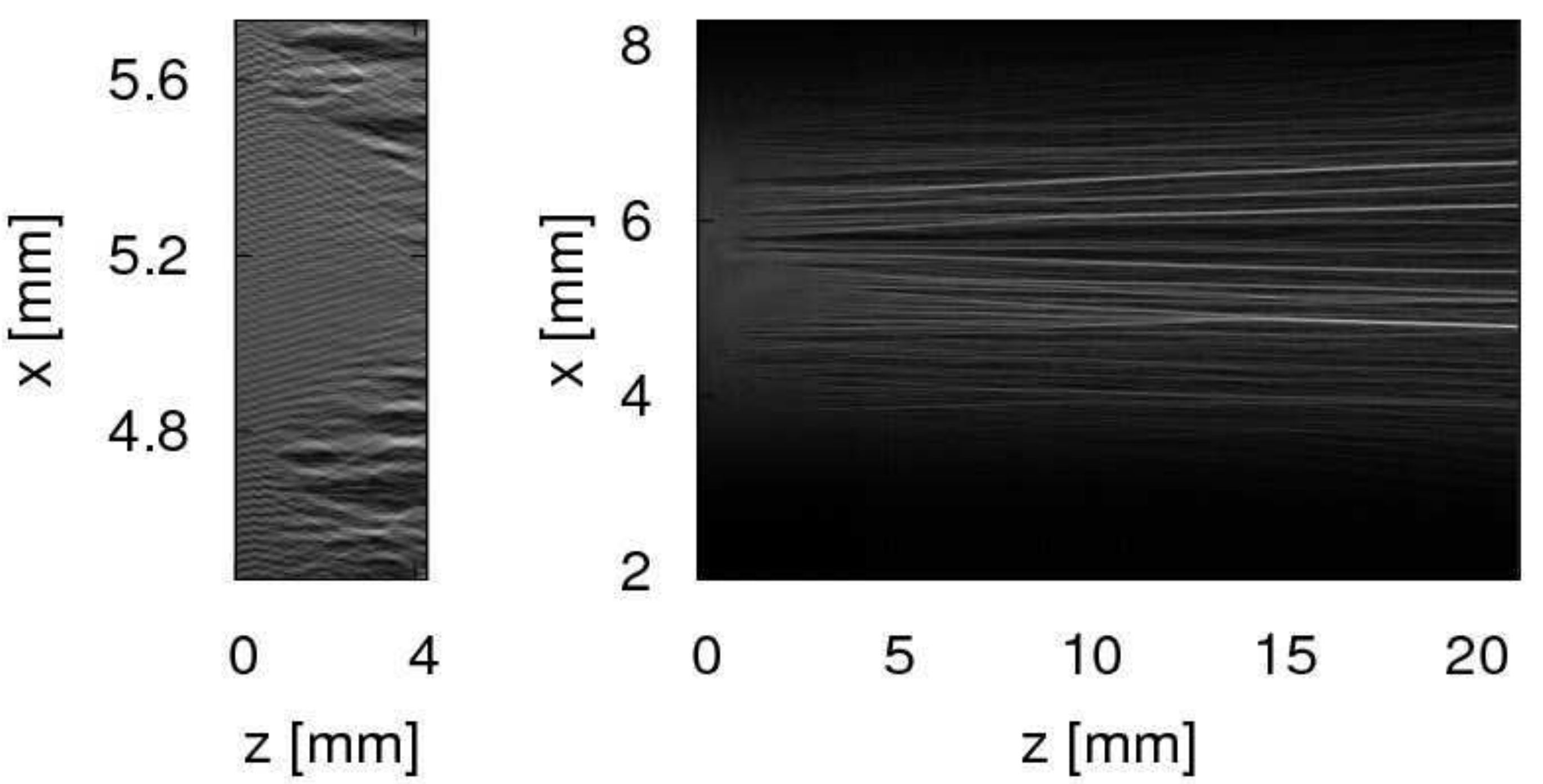}}
\caption{Numerical results for intensity distribution $I(x,z)$. The frame on the
left is a magnified section of the initial propagation of the beam.}
\label{fig:xz-num}
\end{figure}

\begin{figure}[htbp]
\centerline{
\includegraphics[width=8.3cm]{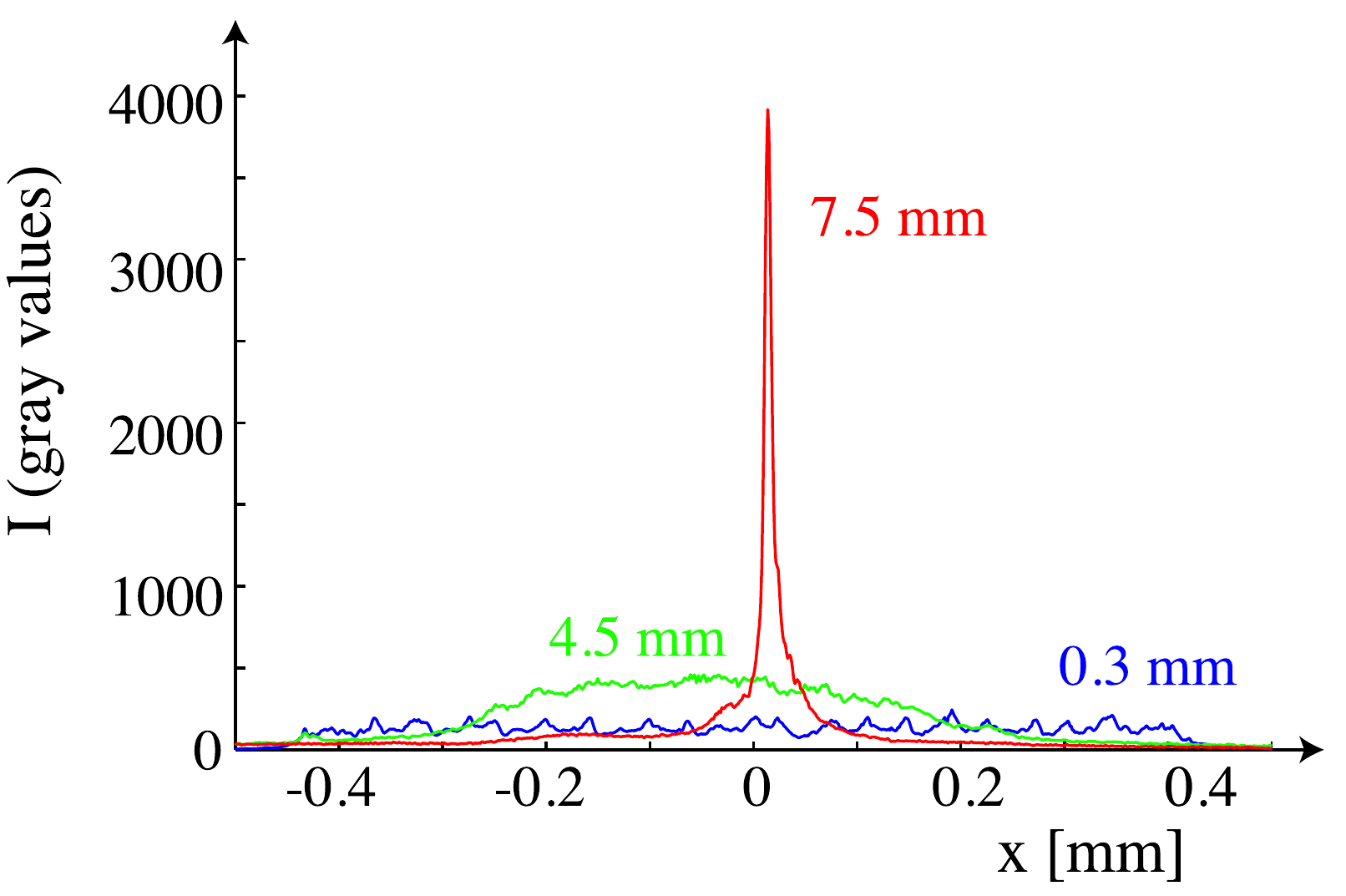}}
\caption{(color online) Linear intensity profiles $I(x)$ taken at different
propagation distances, $z=0.3$, $4.5$ and $7.5$ $mm$.}
\label{fig:xz-profile}
\end{figure}

Example zooms of the intensity distribution $I(x,z)$ showing the beam evolution
during propagation in the experiment and in the numerics are displayed in
Fig.~\ref{fig:xz-exp} and Fig.\ref{fig:xz-num} respectively.
In the high resolution insets on Figs.~\ref{fig:xz-exp} and \ref{fig:xz-num} one
can visually observe that the typical scale increases along the beam which
corresponds to an inverse cascade process. Furthermore, in both
Fig.~\ref{fig:xz-exp} and Fig.~\ref{fig:xz-num} one can see the formation of
coherent solitons out of the random initial field, with
the overall number of solitons reducing as the beam propagates.  Experimentally,
the condensation into one dominant soliton is well revealed by
the linear intensity profiles $I(x)$ taken at different propagation distances,
as shown in Fig.~\ref{fig:xz-profile} for $z=0.3$, $4.5$ and $7.5$ $mm$. Note
that the amplitude of the final dominant soliton is three orders of magnitude
larger than the amplitude of the initial periodic modulation.

\begin{figure}[htbp]
\centerline{
\includegraphics[width=8.3cm]{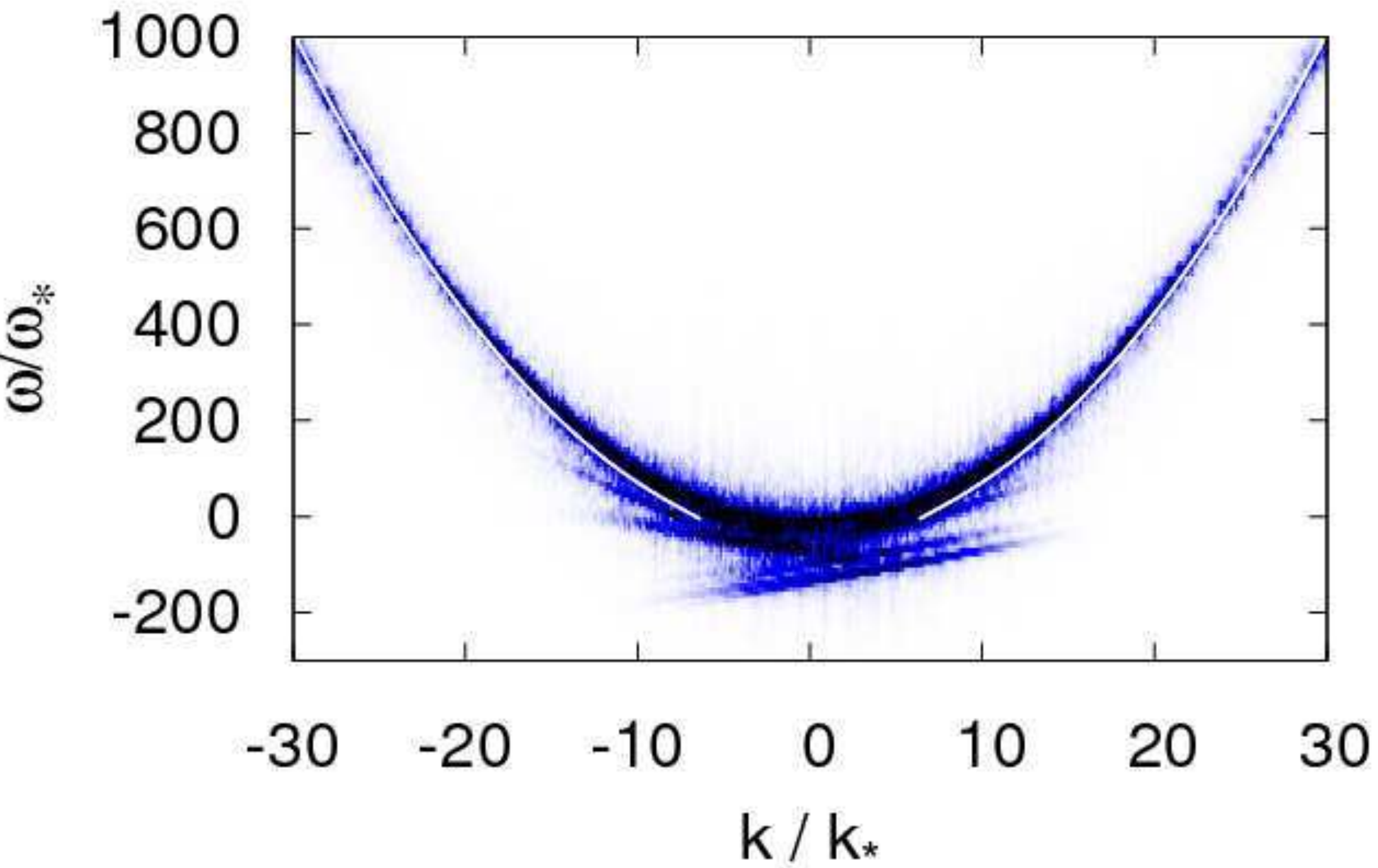}}
\caption{(color online). The $k-\omega$ spectrum of the wave field at $z=2.1$
$m$. $\omega_*=1/256 ql^2_{\xi}$ and $k_* = 1/\sqrt{128} l_{\xi}$. The
Bogoliubov dispersion relation is shown by the solid line.}
\label{fig:wk.pdf}
\end{figure}

We perform two numerical simulations, one with a domain comparable with the
experiment (Figs.~\ref{fig:xz-num} and ~\ref{fig:inverse-num}) and another with
a lower intensity initial condition over a longer propagation distance $z$, to
allow us to clearly distinguish between the WT inverse cascade and MI
(Fig.~\ref{fig:wk.pdf}).
Deviations from integrability result in the interactions of the solitons with
each other and with the random waves, which lead to changes of the soliton
strengths. Both the experimental and the
numerical results in Fig.~\ref{fig:xz-exp} and Fig.~\ref{fig:xz-num}
indicate that solitons gradually merge, reducing the total number.  The observed
increase of the scale and formation of coherent structures represent the
condensation of light.

Separating the random wave and the coherent soliton components can be done
via performing an additional Fourier transform with respect to ``time" $z$ over
a finite $z$-window determined by the frequency scale \cite{NO}. Such
numerically obtained $(k, \omega)$-plot is shown in Fig.~\ref{fig:wk.pdf}.
There, the incoherent
wave component is distributed around the wave dispersion relation, which is
Bogoliubov-modified by the condensate \cite{us} and is shown by a solid line in
Fig.~\ref{fig:wk.pdf} (the line ends at a finite $k$ below which $\omega$
becomes imaginary corresponding to modulational instability). This distribution
is narrow for large $k$ which corresponds to weak nonlinearity, and it gets
wider
toward low $k$, which coincides with the growth of the nonlinearity and the
breakdown of the WT applicability conditions. For these low
$k$ values one can see pieces of slanted
lines (under the dispersion curve) each corresponding to a coherent soliton,
whose speed is equal to the slope.

In conclusion, we have presented the first experimental implementation,
numerical simulations and theory of 1D OWT.
We observe an inverse cascade of photons toward the states with lower
frequencies with a respective KZ spectrum predicted
by the WT theory. The inverse cascade is accompanied by increasing nonlinearity
and the eventual breakdown of WT, leading to
the light condensing into coherent structures - solitons.

This work has been partially supported by the Royal Society's International
Joint Project grant, and by the ANR-07-BLAN-0246-03, {\em turbonde}. U.B.  and
S.R. acknowledge
helpful discussion with Gaetano Assanto and Armando Piccardi.

\end{document}